\documentclass{article}

\usepackage[preprint]{neurips_2026}

\usepackage[utf8]{inputenc} 
\usepackage[T1]{fontenc}    
\usepackage{hyperref}       
\usepackage{url}            
\usepackage{booktabs}       
\usepackage{amsfonts}       
\usepackage{nicefrac}       
\usepackage{microtype}      
\usepackage{xcolor}         
\usepackage{graphicx}
\usepackage{booktabs}
\usepackage{amsmath}
\title{Compact Latent Manifold Translation: A Parameter-Efficient Foundation Model for Cross-Modal and Cross-Frequency Physiological Signal Synthesis}

\author{%
  Bo Cui
  Department of Biomedical Signals and Systems\\
  University of Twente\\
  \texttt{m.r.cui@utwente.nl} \\
  \And
  Xiaowen Song
  Department of Biomedical Signals and Systems\\
  University of Twente\\
  \texttt{x.song@utwente.nl} \\
  \AND
  Yaowen Zhang
  Department of Biomedical Signals and Systems\\
  University of Twente\\
  \texttt{y.zhang-12@utwente.nl}\\
  \AND
  Shunzhe Zhang
  Department of Applied Mathematics\\
  University of Twente\\
  \texttt{s.zhang-9@student.utwente.nl}\\
  \And
  B.J.F. van Beijnum
  Department of Biomedical Signals and Systems\\
  University of Twente\\
  \texttt{b.j.f.vanbeijnum@utwente.nl}\\
  \And
  Monique Tabak
  Department of Biomedical Signals and Systems\\
  University of Twente\\
  \texttt{m.tabak@utwente.nl}\\
  \And
  Ying Wang
  Department of Biomedical Signals and Systems\\
  University of Twente\\
  \texttt{ying.wang@utwente.nl}\\
}

\begin{document}

\maketitle

\begin{abstract}
  The analysis of physiological time series, such as electrocardiograms (ECG) and photoplethysmograms (PPG), is persistently hindered by modality and frequency gaps stemming from heterogeneous recording devices. Existing foundation models typically rely on continuous latent spaces, which frequently suffer from severe modality entanglement, lack high-fidelity cross-frequency generative capacity, and impose high computational costs that prohibit edge-device deployment. In this paper, we propose Compact Latent Manifold Translation (CLMT), a highly parameter-efficient (0.09B) unified framework that bridges these gaps through a novel two-stage discrete translation paradigm. First, we introduce a Universal Tokenizer utilizing Hierarchical Residual Vector Quantization (RVQ) to decouple heterogeneous signals into isolated, well-structured discrete latent manifolds, effectively preventing inter-modality interference. Second, a Context-Prompted Latent Translator maps these discrete tokens across modalities by integrating static physiological priors, reframing complex signal synthesis as a pure latent sequence translation task. Extensive evaluations demonstrate that our 0.09B model significantly outperforms massive baselines. In cross-modal PPG-to-ECG synthesis, it resolves temporal phase drift and dramatically improves the clinical R-peak detection F1-score from 0.37 (baseline) to 0.83. Furthermore, in extreme cross-frequency super-resolution (25Hz to 100Hz), it successfully recovers high-frequency diagnostic landmarks, achieving an unprecedented Pearson correlation of 0.9956. By learning a universal discrete language for biological signals with a fraction of the computational footprint, our approach sets a new trajectory for edge-deployable, multi-modal medical foundation models.

\end{abstract}
\vspace{-1.5em}
\section{Introduction}
\vspace{-1em}
The proliferation of wearable technology has transformed continuous physiological monitoring into a pillar of digital health \cite{dunn2018wearables}. However, the inherent heterogeneity of biosignal data remains a fundamental barrier to developing robust foundation models \cite{csfm2026, bent2020biomarkers}. Clinical devices (e.g., 250--500Hz ECG \cite{zhang2023maefe}) and consumer wearables (e.g., 50Hz PPG \cite{ppgflowecg2025}) present vast physical and temporal discrepancies. Unifying these streams requires effective cross-modal and cross-frequency alignment \cite{physioomni2024}. Traditional sequence processing forces strict temporal alignment via specialized models \cite{csfm2026} or aggressive resampling \cite{ppgflowecg2025}, causing severe temporal distortion \cite{zhou2025mmae} and the irreversible loss of high-frequency diagnostic landmarks (e.g., QRS morphological variations).

Current state-of-the-art (SOTA) foundation models (e.g., CSFM \cite{csfm2026}, PhysioOmni \cite{physioomni2024}) utilize continuous Transformer architectures optimized primarily for representation. They lack generative cross-frequency capacity: their reliance on fixed-size continuous patching structurally prevents synthesizing high-frequency details from low-frequency prompts. Furthermore, continuous joint spaces entangle features of different physical origins, hindering precise cross-modal mapping (e.g., synthesizing clinical-grade ECG from PPG). While specialized generative frameworks like PPGFlowECG \cite{ppgflowecg2025} attempt modality alignment, they remain constrained to single-frequency domains.

To address these challenges, we propose Compact Latent Manifold Translation (CLMT), a unified framework for heterogeneous biosignals. Our core innovation is a Universal Tokenizer based on Hierarchical Residual Vector Quantization (RVQ), which maps multimodal signals into a shared discrete manifold. This hierarchical structure effectively decouples physiological dynamics: top-level quantizers capture shared macro-rhythms across modalities, while bottom-level quantizers preserve sensor-specific high-frequency morphologies. Within this decoupled space, we introduce a \textbf{Context-Prompted Latent Translator} for cross-modal mapping (PPG $\rightarrow$ ECG) and a \textbf{Latent Sequence MAE} for cross-frequency super-resolution (25Hz $\rightarrow$ 100Hz). Crucially, these lightweight modules are trained via continuous feature distillation against a strictly frozen Universal Tokenizer. During inference, routing continuous predictions through the frozen codebook induces a discrete "snapping" effect. This forcefully aligns outputs with established physiological priors, completely circumventing the regression-to-the-mean blurring inherent to continuous generative models.

Our core contributions are summarized as follows: 
\textbf{(1) Representation Paradigm Shift:} We propose a Universal Tokenizer that projects heterogeneous sampling rates and modalities into a hierarchical discrete latent space, eliminating the dependence on resampling-based alignment. 
\textbf{(2) Hierarchical Decoupling Mechanism:} Through Hierarchical RVQ, we achieve multi-scale separation of macro-rhythms and micro-morphologies, preventing feature entanglement and ensuring clear manifold isolation. 
\textbf{(3) Exceptional Efficiency and Snapping Generation:} By freezing the Stage 1 Tokenizer, downstream adaptations require minimal optimization (total framework footprint: 0.09B). The frozen codebook's snapping effect synthesizes sharp clinical landmarks (boosting R-peak F1-score to 0.83), significantly outperforming massive SOTA continuous baselines.

\vspace{-1em}
\section{Related Works}
\vspace{-0.5em}
\subsection{Generative Models for Physiological Time-Series}
\vspace{-0.5em}
While universal time-series models (e.g., TimesNet \cite{timesnet2023}, PatchTST \cite{patchtst2023}) set general benchmarks, the stringent electrophysiological constraints of biosignals necessitate domain-specific foundation paradigms. Currently, developing such models faces a tripartite challenge: high-fidelity cross-modal translation, generative cross-frequency synthesis (super-resolution), and compact edge deployment \cite{csfm2026, Li_2026}. 

The research landscape remains fragmented across these dimensions. Multi-modal architectures like PhysioOmni \cite{physioomni2024} introduce unified representations but lack the generative mechanism for high-frequency morphological synthesis. Conversely, specialized generative frameworks \cite{uap2e2025, ppgflowecg2025, e2ep2e2026} bridge the modality gap but are restricted to single-frequency domains, failing to reconstruct clinical-grade landmarks from low-power wearable sensors. Regarding super-resolution (SR), Multi-scale MAE \cite{zhou2025mmae} focuses primarily on anomaly detection rather than unified cross-manifold generation. Finally, while CSFM \cite{csfm2026} demonstrates validated cross-modal performance, its massive parameter footprint and lack of dedicated SR layers hinder edge deployment. CLMT addresses these gaps through a unified, parameter-efficient discrete manifold.
\vspace{-1em}
\subsection{Discrete Representation and Vector Quantization}
\vspace{-0.5em}
Discrete representation learning, pioneered by VQ-VAE \cite{van2017neural} and advanced by audio codecs \cite{defossez2022high, borsos2023audiolm}, effectively compresses continuous features into discrete codewords. Recently, BioCodec \cite{avramidis2025biocodec} and BrainRVQ \cite{brainrvq2026} successfully extended Residual Vector Quantization (RVQ) to neurophysiological data to capture multi-scale dynamics. However, existing biosignal tokenizers predominantly rely on single-level quantization, which fails to capture the multi-scale, cross-modal properties of heterogeneous signals. 

CLMT distinguishes itself by utilizing hierarchical RVQ to explicitly decouple physiological dynamics based on their distinct electrophysiological and hemodynamic origins. Because heterogeneous signals (ECG and PPG) share a fundamental cardiac cycle but diverge in sensor-specific micro-morphologies, our top-level quantizers encode the shared macro-rhythm, while subsequent layers capture modality-specific high-frequency features. This hierarchical decoupling establishes a unified latent alignment, preserving fundamental rhythms while structurally preparing the space for high-fidelity, modality-specific synthesis.
\vspace{-1em}
\subsection{Masked Modeling in Time Series}
\vspace{-0.5em}
Masked Autoencoding (MAE) has successfully transitioned from computer vision \cite{he2022mae, vaswani2017attention} to ECG analysis \cite{zhang2023maefe, yang2022mae_ecg, rhythmbert2026, icml2025dbeta}. However, current biosignal MAE architectures (e.g., MMAE-ECG \cite{zhou2025mmae}) primarily function as multivariate imputation frameworks. They reconstruct masked segments relying on local contextual cues within a continuous latent space. This continuous operation inevitably leads to statistical averaging, causing the loss of sharp, high-frequency morphological landmarks strictly essential for clinical diagnosis. 

CLMT departs from this imputation convention by deploying MAE within a highly decoupled, discrete latent manifold. By completing masked latent sequences and routing the continuous predictions back through the frozen codebook, our model induces a discrete "snapping" effect. This architectural shift transforms the MAE from a context-based interpolator into a powerful generative framework capable of extreme cross-frequency super-resolution (e.g., 25Hz to 100Hz), successfully recovering diagnostic landmarks that continuous-space models inherently blur.

\vspace{-1em}
\section{Method}
\vspace{-0.5em}
\subsection{Overview}
\vspace{-0.5em}
We propose a two-stage framework for modeling physiological time series across modalities and sampling rates. Given an input signal $\mathbf{X}^{(m)} \in \mathbb{R} ^ {C \times T}$, where $m \in \left\{ \mathrm{ECG} , \mathrm{PPG} \right\}$ denotes the signal modality. $C$ is the number of input channels (e.g., $C=1$ in our setting), and $T$ is the number of the temporal samples. In addition, each sample is associated with static participant-specific attributions: $\mathbf{S} \in \mathbb{R}^k$, where $k$ is the dimension of static features (e.g., demographic or clinical variables). Our goal is to learn a shared representation that enables conditional mapping:
\vspace{-0.5em}
\begin{equation}
    \mathbf{X}_{\mathrm{target}}=f(\mathbf{X}_{\mathrm{source}}, \mathbf{S})
\end{equation}

\begin{figure}[t]
  \vspace{-1em}
  \centering
  \includegraphics[width=\linewidth]{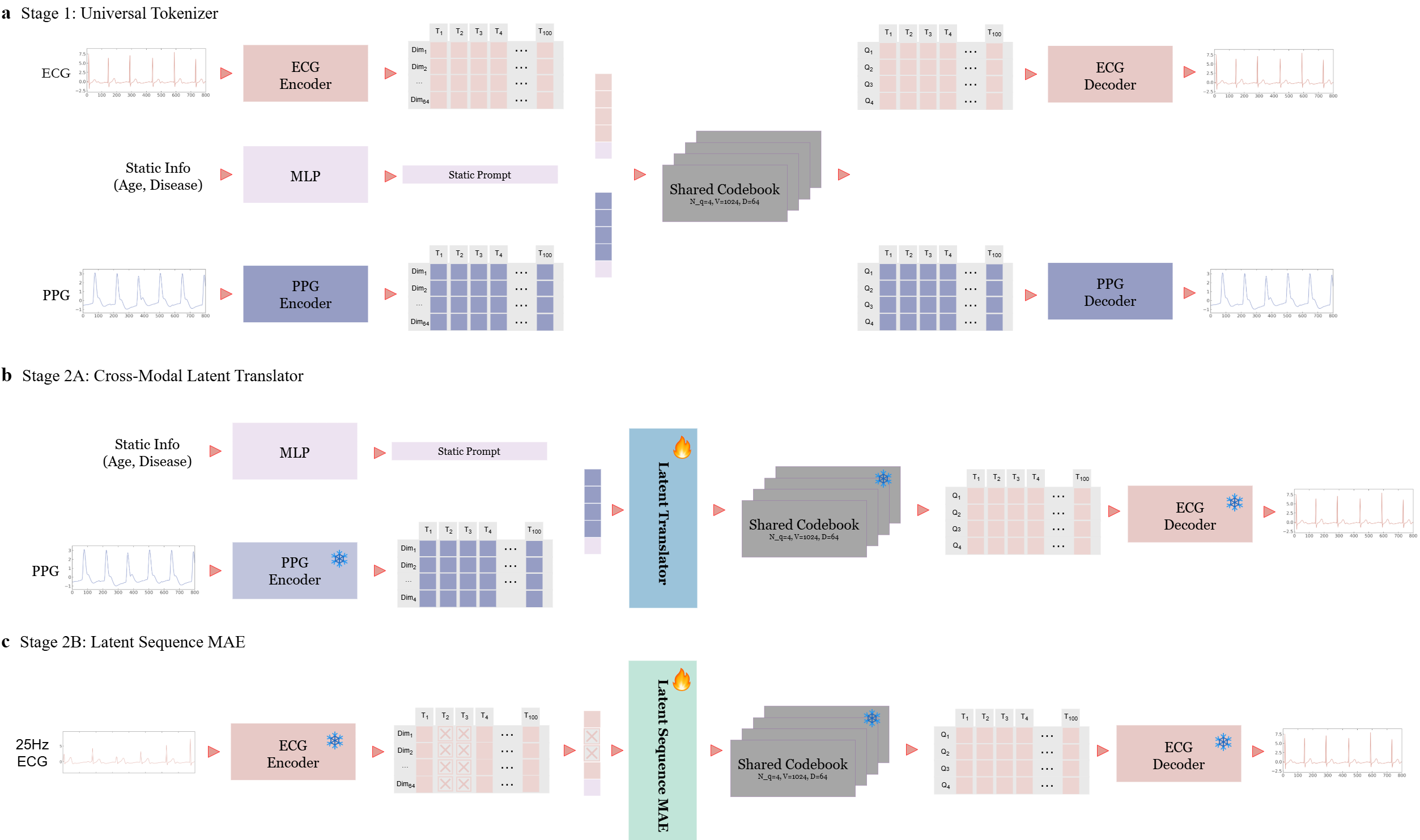} 
  \vspace{-1.5em}
  \caption{
    \textbf{Overall architecture of the Compact Latent Manifold Translation (CLMT) framework.} The framework consists of a foundational tokenizer and two downstream latent tasks. 
    \textbf{(a) Stage 1: Universal Tokenizer.} Heterogeneous physiological signals (ECG and PPG) are encoded, conditioned on static demographic prompts, and mapped into a Shared Hierarchical RVQ Codebook. This stage learns a common discrete physiological representation space across ECG and PPG while preserving modality-specific structural characteristics. 
    \textbf{(b) Stage 2A: Cross-Modal Latent Translator.} For PPG-to-ECG translation, a lightweight Transformer (denoted by the fire icon fire, indicating fine-tuned task-specific weights) maps the source PPG latents to the target ECG continuous space. The predicted features are then snapped to the frozen codebook (denoted by the snowflake icon snowflake, indicating absolute freezing) to restore sharp, high-frequency diagnostic anchors before decoding. 
    \textbf{(c) Stage 2B: Latent Sequence MAE.} For signal super-resolution (25Hz $\rightarrow$ 100Hz), the sparse input results in a masked latent sequence. The Latent Masked Autoencoder (MAE) (denoted by fire) performs completion within the semantic space. By reusing the frozen Stage 1 components (denoted by snowflake), both downstream tasks achieve state-of-the-art generative fidelity with exceptional parameter efficiency (only 0.09B parameters in total).}
  \vspace{-1.2em}
  \label{fig:main_architecture}
\end{figure}

\vspace{-0.5em}
\subsection{Stage 1: Universal Tokenizer}
\vspace{-0.5em}
\subsubsection{Latent Encoding}
\vspace{-0.5em}
We employ a 1D convolutional encoder $E(\cdot )$ to extract latent representations:
\begin{equation}
    \mathbf{Z} = E(\mathbf{X}^{(m)}) \in \mathbb{R}^{N \times d}, 
\end{equation}

where $N$ is the downsampled temporal length and $d$ is the latent feature dimension. The encoder consists of stacked temporal convolutional layers with residual connections. To preserve temporal ordering, we add continuous-time positional encoding to the latent sequence.
\vspace{-0.5em}
\subsubsection{Static Conditioning}
\vspace{-0.5em}
We encode static attributes using a multilayer perceptron (MLP):
\begin{equation}
    \mathbf{P} = f_{\mathrm{prompt}}(\mathbf{S}) \in \mathbb{R}^{d}
\end{equation}
The prompt is broadcast across the sequence: $\tilde{\mathbf{Z}}=\mathbf{Z} + \mathbf{P}$, allowing participant-specific information to modulate all temporal features.
\vspace{-0.5em}
\subsubsection{Hierarchical Residual Vector Quantization (RVQ)}
\vspace{-0.5em}
To discretize the continuous representation, we apply a multi-stage RVQ module. Given the encoder output $\tilde{\mathbf{Z}}$, we initialize the residual as $\mathbf{R}_0 = \tilde{\mathbf{Z}}$, and iteratively quantize it across $L$ stages:
\vspace{-0.5em}
\begin{equation}
    \hat{\mathbf{Z}} = \sum_{l=1}^{L}Q_l (\mathbf{R}_{l-1}), \quad \mathbf{R}_l = \mathbf{R}_{l-1} - Q_l (\mathbf{R}_{l-1})
\end{equation}
\vspace{-0.3em}
where each quantizer $Q_l$ performs a nearest-neighbor lookup within its associated codebook $\mathcal{V}_l=\left\{\mathbf{e}_1^{(l)}, \ldots, \mathbf{e}_K^{(l)}\right\}$. 

Crucially, the hyperparameters $K$ (codebook size) and $L$ (number of layers) provide explicit physiological interpretability. $K$ controls the diversity of discrete physiological patterns captured, while $L$ defines the depth of hierarchical abstraction. Earlier layers ($l=1,2$) encode low-frequency macro-rhythms shared across modalities, whereas deeper layers capture fine-grained, modality-specific high-frequency morphologies. By sharing these codebooks across ECG and PPG, similar underlying physiological states are mapped to identical codewords, enabling implicit cross-modal alignment.
\vspace{-0.5em}
\subsubsection{Modality-specific Decoding}
\vspace{-0.5em}
We reconstruct the waveform using modality-specific decoders: $\hat{\mathbf{X}}^{(m)} = D^{(m)}(\hat{\mathbf{Z}})$, where each decoder is implemented as a convolutional upsampling network with residual blocks, restoring the temporal resolution of the original signal.
\vspace{-0.5em}
\subsubsection{Training Objective}
\vspace{-0.5em}
\label{sec:stage1}
The tokenizer is trained with:
\begin{equation}
    \mathcal{L}_{\mathrm{stage1}} = \mathcal{L}_{\mathrm{recon}} + \lambda_{\mathrm{vq}}\mathcal{L}_{\mathrm{vq}}
\end{equation}
where $\mathcal{L}_{\mathrm{vq}}$ is the standard commitment loss used to align the continuous encoder outputs with the discrete codebook embeddings. To explicitly prevent the attenuation of high-frequency landmarks during this initial space construction, we employ a shape-aware reconstruction loss:
\begin{equation}
\mathcal{L}_{\mathrm{recon}} = \|\hat{\mathbf{X}}-\mathbf{X}\|_1 + \alpha(1-\cos (\hat{\mathbf{X}}, \mathbf{X})) + \beta\|\nabla \hat{\mathbf{X}}-\nabla \mathbf{X}\|_1
\end{equation}
which jointly preserves pointwise amplitude, global phase shape, and local temporal gradient variations (with $\alpha=0.5, \beta=0.5$ set empirically). 

We emphasize that this shape-aware loss is only used to establish the Stage 1 discrete dictionary. In Stage 2 downstream tasks, the baseline models and our adapters are optimized using strictly pure MSE loss. Therefore, the superior sharpness generated by CLMT downstream is not a byproduct of the loss function, but stems inherently from the topological constraint of our discrete codebook.
\vspace{-0.5em}
\subsection{Stage 2A: Downstream Cross-Modal Latent Translation}
\vspace{-0.5em}
\label{sec:stage2a}
We use the frozen encoder $E(\cdot )$ from stage 1 to extract $\mathbf{Z}_{\mathrm{source}} = E(\mathbf{X}_{\mathrm{source}})$. We encode static attribution into a prompt token: $\mathbf{P} \in \mathbb{R}^d$ and the input sequence is constructed as: $\mathbf{Z}_{\mathrm{in}} = [\mathbf{P}, \mathbf{Z}_{\mathrm{source}}]$. We process this sequence using a Transformer encoder:
\begin{equation}
    \mathbf{H} = \mathrm{Transformer}(\mathbf{Z}_{\mathrm{in}}),
\end{equation}
where the Transformer consists of multiple self-attention layers with feedforward networks, enabling global temporal reasoning. We then remove the prompt token and map the remaining sequence:
\begin{equation}
    \mathbf{\hat{Z}}_{\mathrm{target}} = g(\mathbf{H}_{2:(N+1)}),
\end{equation}
where $g(\cdot)$ is a linear projection head. The Transformer is trained with:
\begin{equation}
    \mathcal{L}_{\mathrm{trans}} = \left\| \mathbf{\hat{Z}}_{\mathrm{target}} - \mathbf{Z}_{\mathrm{target}} \right\|_2^2
\end{equation}
\vspace{-1em}
\subsection{Stage 2B: Downstream Latent Sequence MAE for Super-Resolution}
\vspace{-0.5em}
\label{sec:stage2b}
Note that the Stage 1 Universal Tokenizer is pre-trained on high-resolution signals (e.g., 100Hz) to inherently learn a robust high-frequency vocabulary. To evaluate super-resolution, we construct a sparse low-resolution input via temporal subsampling: 
\begin{equation}
    \mathbf{X}_{\mathrm{low}} = \mathrm{Downsample}(\mathbf{X}_{\mathrm{high}}).
\end{equation}
We extract the corresponding continuous latent sequences $\mathbf{Z}_{\mathrm{low}}$ and $\mathbf{Z}_{\mathrm{high}}$ using the frozen encoder $E(\cdot)$. We use a latent reconstruction network $f_{\mathrm{MAE}}$, implemented as a sequence model, to estimate: $\mathbf{\hat{Z}}_{\mathrm{high}}=f_{\mathrm{MAE}}(\mathbf{Z}_{\mathrm{low}})$, with the following training strategy:
\begin{equation}
    \mathcal{L}_{\mathrm{freq}} = \left\| \mathbf{\hat{Z}}_{\mathrm{high} } -  \mathbf{Z}_{\mathrm{high}} \right\|_2^2.
\end{equation}
\vspace{-1em}
\subsection{Training Strategy}
\vspace{-0.5em}
All models are trained on NVIDIA RTX A6000 Pro GPUs using the AdamW optimizer \cite{adam} with an initial learning rate of $1 \times 10^{-3}$, weight decay of $1 \times 10^{-4}$, and a batch size of 64. Early stopping is applied with a patience of 5 epochs based on validation loss.

\textbf{Stage 1 (Universal Tokenizer).}
The encoder-decoder backbone and hierarchical RVQ codebook are trained jointly on heterogeneous modalities (ECG and PPG) using the shape-aware objective and VQ commitment loss described in Section \ref{sec:stage1}.

\textbf{Stage 2 (Downstream Latent Operations via Continuous Distillation).}
For both the Latent Translator (Section \ref{sec:stage2a}) and the Cross-Frequency MAE (Section \ref{sec:stage2b}), the Stage 1 Universal Tokenizer is kept strictly frozen throughout training, including batch normalization statistics. Each Stage 2 network is optimized independently using the MSE loss defined in its respective section. 

Crucially, rather than optimizing these downstream networks in the waveform domain, we frame both tasks as pure latent continuous feature distillation. The target is defined as the continuous latent representation extracted by the frozen Stage 1 encoder. During inference, the continuous predictions ($\mathbf{\hat{Z}}_{\mathrm{target}}$ and $\mathbf{\hat{Z}}_{\mathrm{high}}$) are subsequently routed through the frozen RVQ codebook. This routing enforces a discrete snapping effect, matching the continuous prediction to the nearest established physiological prior.
\vspace{-1em}
\section{Results}
\vspace{-0.5em}
\subsection{High-Fidelity Cross-Modal Translation}
\vspace{-0.5em}
\label{sec:results_translation}
Before evaluating generation quality, we first confirmed via UMAP visualizations that our Universal Tokenizer achieves pristine modality isolation without entanglement (see Appendix \ref{sec:results_latent_space} for comprehensive latent space topological analysis).

Translating PPG to ECG is notoriously challenging due to significant morphological gaps and latency variations (e.g., Pulse Transit Time). We evaluate our model against the direct PPG to ECG reconstruction (P2E) baseline, analyzing quantitative fidelity, waveform morphology, and latent temporal dynamics.

\textbf{Quantitative Superiority and Clinical Utility.} 
Table \ref{tab:cross_modal_results} demonstrates our model's statistically significant improvements over the P2E baseline. We reduce RMSE from 0.6805 to \textbf{0.4245} and halve the Fréchet Distance (FD) from 37.49 to \textbf{16.17}, indicating superior distributional fidelity. Most remarkably, the clinical R-peak detection F1-score surges from 0.3724 to \textbf{0.8265}. This proves our framework successfully synthesizes critical high-frequency diagnostic anchors rather than merely approximating low-frequency trends.
\begin{table}[htbp]
  \vspace{-1em} 
  \centering
  \small 
  \setlength{\tabcolsep}{3.5pt} 
  \caption{
    \textbf{Quantitative Evaluation of Cross-Modal Reconstruction.} Our Master Model significantly outperforms the End-to-End baseline across all metrics. The dramatic improvement in the F1-score highlights our model's unique capability to preserve clinically vital high-frequency features (e.g., R-peaks), avoiding the over-smoothing typical of direct regression baselines. (* denotes the best cross-modal performance).
  }
  \label{tab:cross_modal_results}
  \begin{tabular}{l c c c c c}
    \toprule
    \textbf{Metric} & \textbf{E2E (Upper)} & \textbf{P2P (Upper)} & \textbf{Ours (Master)} & \textbf{Baseline (P2E)} & \textbf{w/o Static} \\
    \midrule
    RMSE $\downarrow$  & 0.0725 & 0.0088 & \textbf{0.4245}* & 0.6805 & 0.7127 \\
    MAE $\downarrow$   & 0.0459 & 0.0064 & \textbf{0.2517}* & 0.4760 & 0.4717 \\
    F1 $\uparrow$      & 0.9208 & N/A    & \textbf{0.8265}* & 0.3724 & 0.8032 \\
    FD $\downarrow$    & 0.3267 & 0.0047 & \textbf{16.1779}*& 37.4933& 33.7659 \\
    \bottomrule
  \end{tabular}
  \vspace{-1em} 
\end{table}

\begin{figure}[t]
  \centering
    \includegraphics[width=0.8\linewidth]{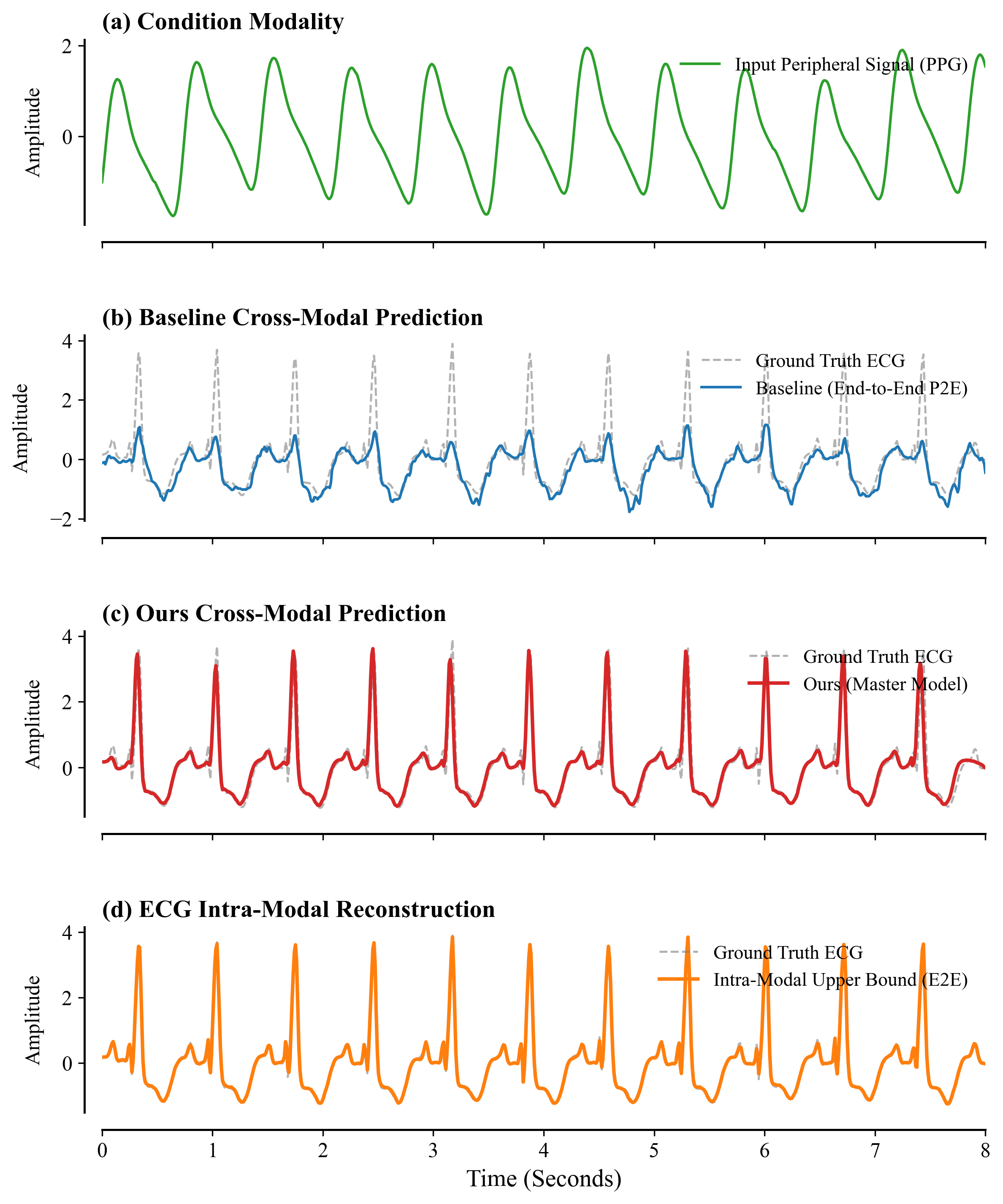}
  \caption{
    \textbf{Waveform qualitative comparison for PPG-to-ECG translation.} \textbf{(a)} The input peripheral condition modality (PPG). \textbf{(b)} The P2E baseline fails to align temporal phases and severely flattens the high-frequency QRS complexes due to regression-to-the-mean. \textbf{(c)} Our Master Model successfully reconstructs sharp, precise diagnostic features, tightly matching the ground truth ECG. \textbf{(d)} The intra-modal ECG reconstruction (i.e., pure ECG-to-ECG autoencoding) serves as the theoretical performance upper bound for the decoder.
  }
  \label{fig:waveform_recon}
  \vspace{-2.5em}
\end{figure}

\textbf{Overcoming Morphological Flattening.} 
These gains are visually evident in Figure \ref{fig:waveform_recon}. The continuous P2E baseline suffers from the "regression-to-the-mean" dilemma, producing severely attenuated and temporally shifted QRS complexes (Figure \ref{fig:waveform_recon}b). Conversely, by routing features through the discrete physiological priors learned in Stage 1, our Latent Translator reconstructs sharp, precise QRS complexes (Figure \ref{fig:waveform_recon}c) that approach the intra-modal upper bound (i.e., pure ECG autoencoding limit).

\textbf{Resolving Temporal Phase Mismatch.} 

\begin{figure}[t]
  \centering
    \includegraphics[width=\linewidth]{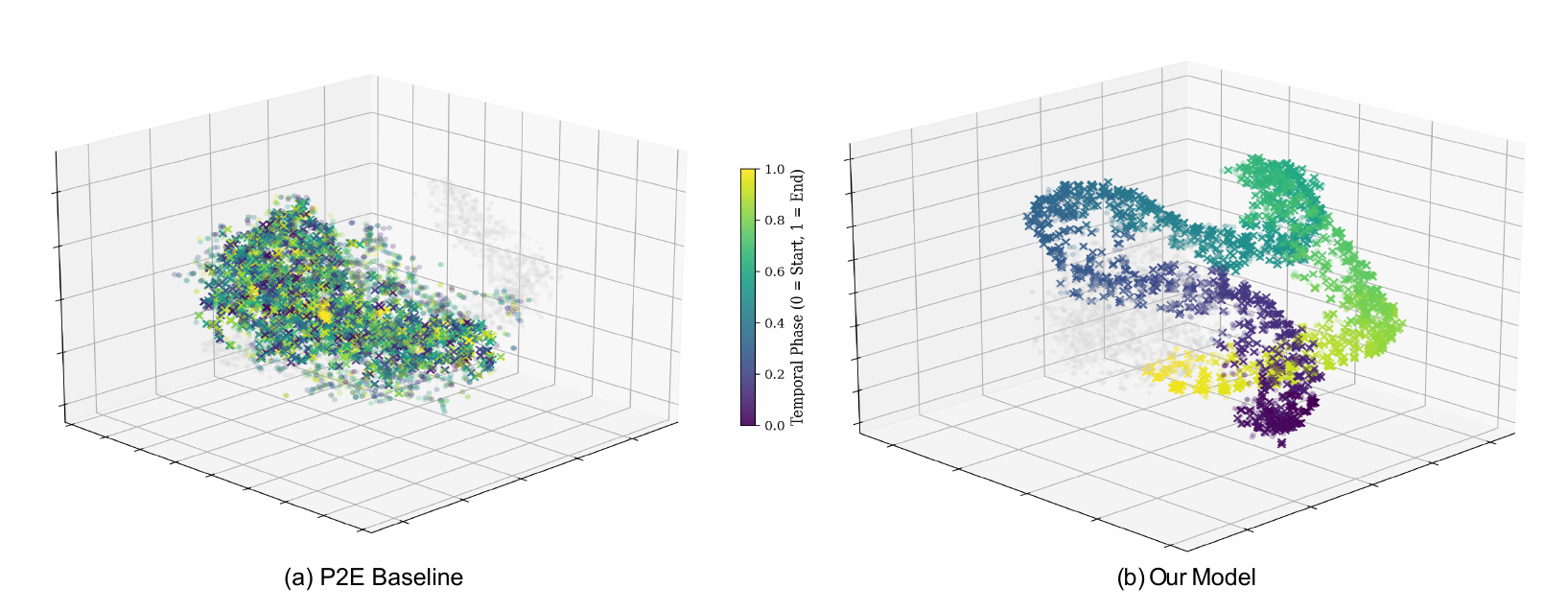}
  \caption{
    \textbf{3D t-SNE of latent temporal dynamics.} Colors encode normalized cardiac phase (0\,=\,onset, 1\,=\,end). \textbf{(a)} The P2E baseline collapses all distributions (PPG input, translated ECG, true ECG) into a phase-entangled cluster. \textbf{(b)} Our model maps translated and ground-truth ECG into a coherent helical manifold with a monotonic phase gradient, while strictly preserving spatial separation from the input PPG embeddings. \textit{Note: 3D axes represent dimensionless t-SNE components.}
  }
  \label{fig:latent_dynamics}
  \vspace{-1em}
\end{figure}

The underlying driver of this performance gap lies in latent temporal dynamics (Figure~\ref{fig:latent_dynamics}). In the P2E baseline (Figure~\ref{fig:latent_dynamics}a), input PPG, translated ECG, and ground-truth ECG embeddings collapse into a single unstructured cluster. This \textit{temporal phase entanglement} conflates cardiac cycle stages, destroying sequential structure. By contrast, our model (Figure~\ref{fig:latent_dynamics}b) strictly segregates the input PPG latents while organizing the translated and ground-truth ECG embeddings into a geometrically coherent \textit{helical manifold}. This smooth, monotonic phase gradient confirms that our Latent Translator performs a genuine cross-modal mapping, implicitly encoding the chronological progression of the cardiac cycle without collapsing modality boundaries.

\vspace{-0.5em}
\subsection{Cross-Frequency Adaptation: MAE-Driven Signal Super-Resolution}
\vspace{-0.5em}
\label{sec:results_sr}
To evaluate generalizability, we assess a 4$\times$ super-resolution (SR) task: upsampling from a sparse 25\,Hz input to a 100\,Hz target. We compare our latent Masked Autoencoder (MAE)---which operates directly on the quantized sequence generated by the frozen Stage 1 Tokenizer---against classical interpolations and a parameter-matched time-domain 1D CNN.

\begin{figure}[t]
  \centering
    \includegraphics[width=\linewidth]{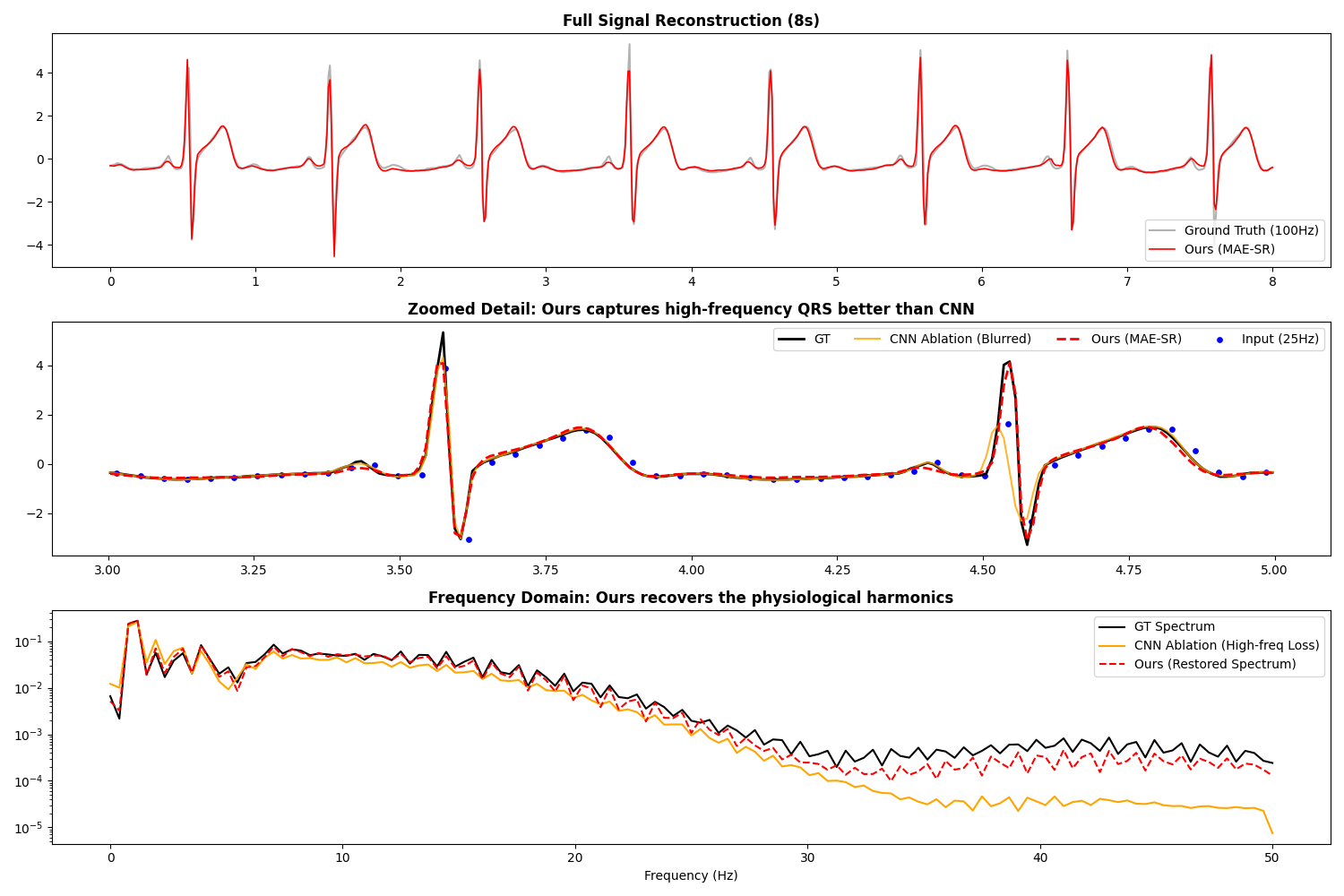}
    \caption{
        \textbf{4X Super-Resolution Evaluation.} \textbf{(Top)} Full 8-second reconstruction securely tracks the 100\,Hz ground truth. \textbf{(Middle)} Zoomed-in QRS complexes highlight the continuous CNN's blurring effect (regression-to-the-mean), whereas our discrete MAE perfectly infers high-frequency spikes from sparse 25\,Hz inputs (blue dots). \textbf{(Bottom)} Spectral analysis confirms our model restores $> 20$\,Hz harmonics lost by the CNN. \textit{Note: Waveforms depict Amplitude (mV) vs. Time (s); Spectra depict Power Density vs. Frequency (Hz).}
    }
  \label{fig:sr_reconstruction}
  \vspace{-1em}
\end{figure}

\textbf{Overcoming Regression-to-the-Mean.}
Time-domain regression models typically suffer from over-smoothing. As shown in Figure~\ref{fig:sr_reconstruction} (middle), the CNN baseline severely attenuates sharp QRS complexes, failing to recover high-frequency structures from sparse 25\,Hz anchors. Conversely, our MAE operates within the discrete latent space. Rather than regressing to a blurred average, it selects from physiologically grounded codebook entries, synthesizing substantially sharper R-peaks. Spectrally (Figure~\ref{fig:sr_reconstruction}, bottom), the CNN loses harmonics above 20\,Hz, whereas our MAE faithfully tracks the full ground-truth power spectrum.

\textbf{Quantitative Evaluation.}
Table~\ref{tab:sr_results} confirms these advantages. Compared to the CNN, our model halves the RMSE and achieves a Pearson Correlation of \textbf{0.9956}. Crucially, in metrics sensitive to morphological and spectral fidelity, our MAE achieves an R-peak F1-score of \textbf{0.9248} and a PSD-RMSE of \textbf{0.0013}. Furthermore, the Fréchet Distance (FD) of \textbf{0.0406} highlights that our discrete prior promotes physiologically plausible distributions rather than merely minimizing pointwise error.
\begin{table}[htbp]
  \vspace{-1em} 
  \centering
  \small 
  \setlength{\tabcolsep}{4pt} 
  \caption{
     \textbf{Quantitative Evaluation of 4X Signal Super-Resolution (25Hz $\rightarrow$ 100Hz).} Our latent MAE significantly outperforms all baselines. The exceptional F1, PSD-RMSE, and FD scores demonstrate an unparalleled ability to recover high-frequency structural details and spectral harmonics. (* denotes best performance).
  }
  \label{tab:sr_results}
  \begin{tabular}{l c c c c}
    \toprule
    \textbf{Metric} & \textbf{Linear} & \textbf{Cubic} & \textbf{CNN} & \textbf{Ours (MAE-SR)} \\
    \midrule
    RMSE $\downarrow$      & 0.3708 & 0.4510 & 0.2053 & \textbf{0.0847}* \\
    MAE $\downarrow$       & 0.1634 & 0.1910 & 0.1028 & \textbf{0.0353}* \\
    PCC $\uparrow$         & 0.9212 & 0.8862 & 0.9773 & \textbf{0.9956}* \\
    F1 $\uparrow$          & 0.7231 & 0.7529 & 0.8349 & \textbf{0.9248}* \\
    PSD-RMSE $\downarrow$  & 0.0138 & 0.0078 & 0.0039 & \textbf{0.0013}* \\
    FD (macro) $\downarrow$& 11.7139& 15.0424& 1.3209 & \textbf{0.0406}* \\
    \bottomrule
  \end{tabular}
  \vspace{-1.5em} 
\end{table}

\vspace{-0.5em}
\subsection{Ablation Studies on Architectural Dynamics}
\vspace{-0.5em}
\label{sec:results_ablation}
To isolate component contributions, we ablate our two-stage translation pipeline and the static prompt mechanism on the challenging cross-modal task (Table~\ref{tab:cross_modal_results}).

\textbf{Necessity of the Two-Stage Translation.}
Removing the Stage 2 Translator and reverting to continuous end-to-end regression (\textbf{Baseline E2E-P2E}, bypassing the codebook) catastrophically degrades the R-peak F1-score from $0.8265$ to $0.3724$. This stark contrast proves that cross-modal generation requires two complementary mechanisms: the Stage 1 Tokenizer establishes the indispensable morphological vocabulary (resolving spatial heterogeneity), while the Stage 2 sequence-to-sequence alignment is strictly necessary to resolve inter-modality temporal offsets (e.g., Pulse Transit Time).

\textbf{Role of Static Prompts as Morphological Conditioning.}
Ablating patient-specific metadata (\textbf{Ablation: No Static}) reveals an informative functional dissociation. Temporal phase alignment is largely preserved independently (F1 decreases modestly to $0.8032$), but morphological fidelity degrades substantially: RMSE rises to $0.7127$ and Fréchet Distance doubles to $33.77$. This confirms a structural specialization: the Latent Translator governs temporal rhythm, while static prompts condition subject-specific amplitude scaling and waveform shaping. Without this personalization, the model regresses to a population-average temporal structure that fails to reflect individualized clinical variations.

\vspace{-1em}
\section{Discussion and Conclusion}
\label{sec:discussion}
\vspace{-0.5em}
\subsection{Decoding the Cardiovascular Manifold in Latent Space}
\vspace{-0.5em}
Historically, physiological models were modality-siloed \cite{Guarrasi_2025}. While recent foundation models (e.g., CSFM \cite{csfm2026}) pioneered integration, they typically rely on continuous contrastive alignment, inadvertently forcing \textit{modality entanglement} that homogenizes distinct morphological priors. 

In contrast, CLMT harmonizes multimodal dynamics without conflation. UMAP projections confirm our Universal Tokenizer achieves strict \textit{modality isolation}—mapping ECG and PPG into a shared discrete vocabulary while preserving distinct topological boundaries (representing electrical and mechanical origins, respectively). Building upon this pristine space, t-SNE visualizations (Figure \ref{fig:latent_dynamics}) demonstrate that the Stage 2 Latent Translator successfully decodes the fundamental \textit{cardiac cycle}. By routing isolated PPG latents into the target ECG manifold to form a continuous, phase-aligned helical structure, our model explicitly resolves the inherent phase delay (Pulse Transit Time) between modalities. Ultimately, CLMT intrinsically reverse-engineers the electro-mechanical coupling of the cardiovascular system rather than merely memorizing waveforms.
\vspace{-0.5em}
\subsection{Clinical Relevance and Computational Efficiency}
\vspace{-0.5em}
Beyond addressing practical clinical scenarios—such as zero-shot ECG estimation from wearables and historical data super-resolution—CLMT demonstrates exceptional computational efficiency. Compared to continuous baselines like CSFM-Base (117M parameters) \cite{csfm2026}, our entire framework requires only 90M (0.09B) parameters (a 23\% reduction). Despite this compact footprint, it strictly outperforms massive models in high-frequency feature restoration (e.g., R-peak F1-score of 0.8265). This quantized efficiency overcomes severe resource constraints \cite{Li_2026}, bringing high-fidelity physiological generation closer to real-time edge deployment.
\vspace{-0.5em}
\subsection{Limitations and Future Directions}
\vspace{-0.5em}
We identify two limitations prior to deployment. \textbf{1) Sensitivity to Out-of-Distribution Pathologies:} For atypical arrhythmias absent from the training distribution, RVQ quantization may map anomalous patterns to nearest-neighbor tokens, potentially regularizing clinically significant irregular dynamics \cite{11435839}. \textbf{2) Dependence on Static Conditioning:} In data-scarce emergencies lacking patient metadata, the model defaults to population-average waveforms \cite{clef2024,article1}, limiting precision medicine utility.

In the future, We aim to scale the pre-training corpus to encompass rare pathologies, enriching the discrete vocabulary. Furthermore, the highly structured nature of our latent space offers a promising interpretability avenue: future work will investigate mapping specific codebook sequences to established disease phenotypes, potentially serving as novel, computationally efficient digital biomarkers.
\vspace{-0.5em}
\subsection{Conclusion}
\vspace{-0.5em}
We present CLMT, a two-stage discrete latent framework for physiological signal generation that resolves the temporal phase misalignment and over-smoothing limitations inherent to continuous end-to-end approaches. By establishing a shared discrete codebook that preserves modality-specific structural integrity, our architecture allows downstream modules to exclusively focus on sequence-level semantic alignment. Experimental results across cross-modal translation (PPG-to-ECG) and extreme cross-frequency super-resolution (25Hz $\rightarrow$ 100Hz) demonstrate unparalleled waveform fidelity and clinical landmark preservation. By unifying heterogeneous biological signals into a compact, decipherable language, this work sets a highly efficient and interpretable foundation for the next generation of multimodal medical AI.
\newpage
\appendix
\section{Universal Feature Alignment in Discrete Latent Space}
\label{sec:results_latent_space}

\begin{figure}[t]
  \centering
  \begin{minipage}{0.45\linewidth}
    \centering
    \includegraphics[width=\linewidth]{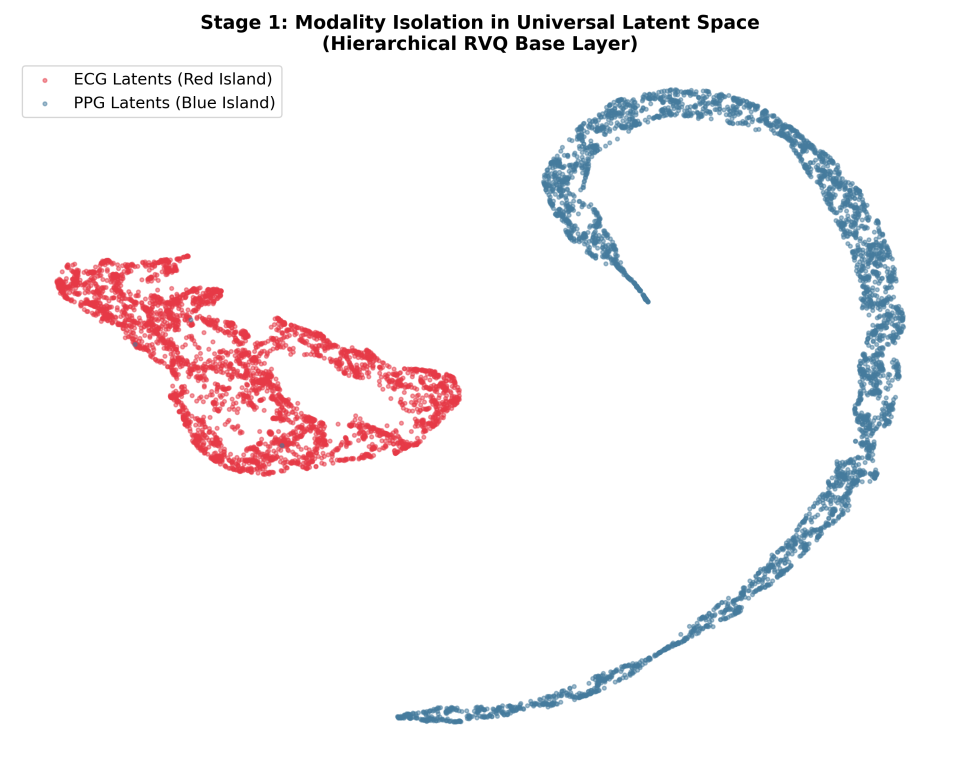}
    \vspace{0.5em}
    
    (a) Our Universal Tokenizer
  \end{minipage}
  \hfill
  \begin{minipage}{0.45\linewidth}
    \centering
    \includegraphics[width=\linewidth]{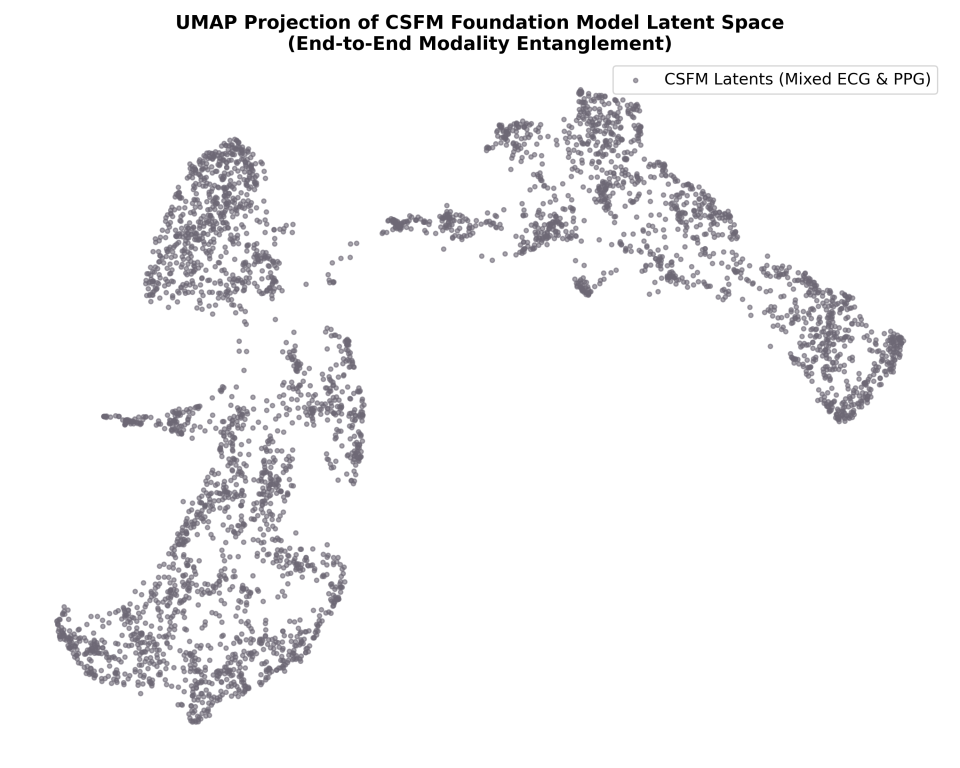}
    \vspace{0.5em}
    
    (b) CSFM Baseline \cite{csfm2026}
  \end{minipage}

  \caption{
    \textbf{UMAP visualizations of the latent space comparing our Universal Tokenizer and the CSFM baseline.} 
    \textbf{(a)} Our model achieves \textit{structured modality isolation}: despite sharing the same discrete RVQ codebook, ECG (Red) and PPG (Blue) embeddings organically form distinct, highly organized sub-manifolds that preserve their intrinsic physical dynamics. 
    \textbf{(b)} In contrast, the CSFM baseline suffers from \textit{modality entanglement}, exhibiting overlapping distributions with blurred boundaries between the signals. The pristine topological separation in our shared space provides a crucial prerequisite for precise, ambiguity-free cross-modal translation in Stage 2.
  }
  \label{fig:latent_umap}
\end{figure}

The foundation of our framework relies on the representational quality of the latent space learned by the Universal Tokenizer. An effective physiological foundation model should balance two inherently competing demands: projecting heterogeneous multimodal signals onto a shared representational space for cross-modal generalization, while retaining the distinctive morphological structure of each modality to prevent inter-modal interference \cite{result1}. To empirically validate whether our architecture achieves this balance, we visualize the learned latent manifolds alongside those of the baseline CSFM using UMAP, as shown in Figure~\ref{fig:latent_umap}.

\textbf{Modality Entanglement in the Baseline.}
As shown in Figure~\ref{fig:latent_umap}(b), the latent space of CSFM reflects a globally aligned representation strategy, in which embeddings from heterogeneous physiological signals overlap with prominent bridging regions. While this design choice is well-suited for tasks that benefit from modality-invariant representations, such as cross-modal retrieval or semantic classification, the lack of strict topological boundaries between modalities introduces ambiguity. For fine-grained waveform generation and modality-to-modality translation, it becomes difficult to preserve modality-specific morphological cues—such as the sharp ventricular depolarization events in ECG or the smooth pulsatile envelope in PPG—that are strictly essential for high-fidelity reconstruction.

\textbf{Structured Modality Separation in a Shared Discrete Space.}
In contrast, Figure~\ref{fig:latent_umap}(a) shows the latent space produced by the Hierarchical RVQ base layer of our Universal Tokenizer. Despite mapping both ECG and PPG signals onto the same shared discrete codebook, the learned embeddings self-organize into two geometrically distinct sub-manifolds: a compact, cyclically folded cluster for ECG and a smooth, arc-shaped distribution for PPG. This emergent \textit{modality isolation} arises naturally without applying any explicit contrastive or domain-separation loss functions, indicating that the shared codebook captures modality-intrinsic dynamics rather than imposing a forced, homogenized representation.

This structural separation has a direct implication for our Stage 2 Latent Translator: the well-defined geometry of each sub-manifold provides unambiguous source and target regions for cross-modal routing, significantly reducing the risk of semantic confusion during translation.
\newpage
\section*{NeurIPS Paper Checklist}

\begin{enumerate}

\item {\bf Claims}
    \item[] Question: Do the main claims made in the abstract and introduction accurately reflect the paper's contributions and scope?
    \item[] Answer: \answerYes{} 
    \item[] Justification: The abstract and Introduction (Section 1) explicitly state our contributions, including the hierarchical decoupling mechanism and parameter efficiency, which are empirically validated in the Results (Section 4).

\item {\bf Limitations}
    \item[] Question: Does the paper discuss the limitations of the work performed by the authors?
    \item[] Answer: \answerYes{} 
    \item[] Justification: We explicitly discuss the limitations of our framework, including its sensitivity to out-of-distribution pathologies and dependence on static conditioning, in Section 5.3 (Limitations and Future Directions).

\item {\bf Theory assumptions and proofs}
    \item[] Question: For each theoretical result, does the paper provide the full set of assumptions and a complete (and correct) proof?
    \item[] Answer: \answerNA{} 
    \item[] Justification: This paper introduces an empirical deep learning architecture for physiological signal generation and does not contain formal mathematical theorems or proofs.

    \item {\bf Experimental result reproducibility}
    \item[] Question: Does the paper fully disclose all the information needed to reproduce the main experimental results of the paper to the extent that it affects the main claims and/or conclusions of the paper (regardless of whether the code and data are provided or not)?
    \item[] Answer: \answerYes{} 
    \item[] Justification: The model architecture (Section 3), hyperparameters, objective functions, and training strategies (Section 3.5) are fully detailed in the core paper to ensure reproducibility.

\item {\bf Open access to data and code}
    \item[] Question: Does the paper provide open access to the data and code, with sufficient instructions to faithfully reproduce the main experimental results, as described in supplemental material?
    \item[] Answer: \answerYes{} 
    \item[] Justification: We include the anonymized source code and scripts required to reproduce our models and experiments in the supplemental material.

\item {\bf Experimental setting/details}
    \item[] Question: Does the paper specify all the training and test details (e.g., data splits, hyperparameters, how they were chosen, type of optimizer) necessary to understand the results?
    \item[] Answer: \answerYes{} 
    \item[] Justification: We clearly specify the optimizer (AdamW), learning rates, weight decay, batch sizes, and early stopping patience directly in Section 3.5.

\item {\bf Experiment statistical significance}
    \item[] Question: Does the paper report error bars suitably and correctly defined or other appropriate information about the statistical significance of the experiments?
    \item[] Answer: \answerNo{} 
    \item[] Justification: Due to the high computational cost of retraining the complete two-stage foundation architecture across multiple random seeds, we report single-run deterministic evaluations. However, the performance margins between our framework and the baselines are exceptionally large (e.g., F1-score 0.8265 vs 0.3724), firmly establishing empirical superiority without the strict necessity of formal error bars.

\item {\bf Experiments compute resources}
    \item[] Question: For each experiment, does the paper provide sufficient information on the computer resources (type of compute workers, memory, time of execution) needed to reproduce the experiments?
    \item[] Answer: \answerYes{} 
    \item[] Justification: Section 3.5 explicitly states that all model training and evaluations were conducted using NVIDIA RTX A6000 Pro GPUs.
    
\item {\bf Code of ethics}
    \item[] Question: Does the research conducted in the paper conform, in every respect, with the NeurIPS Code of Ethics \url{https://neurips.cc/public/EthicsGuidelines}?
    \item[] Answer: \answerYes{} 
    \item[] Justification: Our research complies with the Code of Ethics, leveraging pre-existing anonymized physiological data without introducing dual-use harm or unethical deployment risks.

\item {\bf Broader impacts}
    \item[] Question: Does the paper discuss both potential positive societal impacts and negative societal impacts of the work performed?
    \item[] Answer: \answerYes{} 
    \item[] Justification: We discuss positive impacts (e.g., accessible edge-device clinical monitoring) in Section 5.2 and acknowledge negative risks (e.g., generation bias in atypical pathologies) in Section 5.3.
    
\item {\bf Safeguards}
    \item[] Question: Does the paper describe safeguards that have been put in place for responsible release of data or models that have a high risk for misuse (e.g., pre-trained language models, image generators, or scraped datasets)?
    \item[] Answer: \answerNA{} 
    \item[] Justification: The framework deals with low-risk physiological time-series waveforms for medical research, which does not pose the high-risk misuse concerns of language or image generative models.

\item {\bf Licenses for existing assets}
    \item[] Question: Are the creators or original owners of assets (e.g., code, data, models), used in the paper, properly credited and are the license and terms of use explicitly mentioned and properly respected?
    \item[] Answer: \answerYes{} 
    \item[] Justification: All baseline architectures and existing datasets are properly cited in the text and bibliography, respecting their respective open-source terms.

\item {\bf New assets}
    \item[] Question: Are new assets introduced in the paper well documented and is the documentation provided alongside the assets?
    \item[] Answer: \answerNA{} 
    \item[] Justification: The paper does not release new datasets; the primary asset is the codebase, which is documented and discussed in Question 5.

\item {\bf Crowdsourcing and research with human subjects}
    \item[] Question: For crowdsourcing experiments and research with human subjects, does the paper include the full text of instructions given to participants and screenshots, if applicable, as well as details about compensation (if any)? 
    \item[] Answer: \answerNA{} 
    \item[] Justification: The research relies exclusively on existing, publicly available, anonymized physiological datasets and does not involve new human subjects or crowdsourcing.

\item {\bf Institutional review board (IRB) approvals or equivalent for research with human subjects}
    \item[] Question: Does the paper describe potential risks incurred by study participants, whether such risks were disclosed to the subjects, and whether Institutional Review Board (IRB) approvals (or an equivalent approval/review based on the requirements of your country or institution) were obtained?
    \item[] Answer: \answerNA{} 
    \item[] Justification: Our work utilizes pre-existing, de-identified public datasets that do not require separate IRB approval for secondary computational analysis.

\item {\bf Declaration of LLM usage}
    \item[] Question: Does the paper describe the usage of LLMs if it is an important, original, or non-standard component of the core methods in this research? Note that if the LLM is used only for writing, editing, or formatting purposes and does \emph{not} impact the core methodology, scientific rigor, or originality of the research, declaration is not required.
    \item[] Answer: \answerNA{} 
    \item[] Justification: Large Language Models were not utilized as an original or methodological component of the research; they were only used for text editing and formatting.

\end{enumerate}
\newpage
\bibliographystyle{unsrtnat}
\bibliography{main}

\end{document}